\documentclass[
reprint,
 amsmath,amssymb,
aps,prb,
]{revtex4-1}

\usepackage{graphicx}
\usepackage{dcolumn}
\usepackage{bm}
\usepackage{textcomp, gensymb}
\usepackage{lmodern}
\usepackage{hyperref}
\newcommand*{\citen}[1]{%
  \begingroup
    \romannumeral-`\x 
    \setcitestyle{numbers}%
    \cite{#1}%
  \endgroup   
}

\begin{document}

\preprint{APS/123-QED}

\title{\textbf{Towards the understanding of the origin of charge-current-induced spin voltage signals in the topological insulator Bi$_2$Se$_3$}}

\author{E. K. de Vries$^1$}
	\email{eric.de.vries@rug.nl}
\author{A. M. Kamerbeek$^1$}
\author{N. Koirala$^2$}
\author{M. Brahlek$^2$}
\author{M. Salehi$^3$}
\author{S. Oh$^2$}
\author{B. J. van Wees$^1$}
\author{T. Banerjee$^1$}
	\email{t.banerjee@rug.nl}
\affiliation{$^1$Zernike Institute for Advanced Materials, University of Groningen, Groningen, The Netherlands}
\affiliation{$^2$Department of Physics $\&$ Astronomy, Rutgers, The State University of New Jersey, Piscataway, New Jersey 08854, U.S.A.}
\affiliation{$^3$Department of Materials Science and Engineering, Rutgers, The State University of New Jersey, Piscataway, New Jersey 08854, U.S.A.}

\begin{abstract}
Topological insulators provide a new platform for spintronics due to the spin texture of the surface states that are topologically robust against elastic backscattering. Here, we report on an investigation of the measured voltage obtained from efforts to electrically probe spin-momentum locking in the topological insulator Bi$_2$Se$_3$ using ferromagnetic contacts. Upon inverting the magnetization of the ferromagnetic contacts, we find a reversal of the measured voltage. Extensive analysis of the bias and temperature dependence of this voltage was done, considering the orientation of the magnetization relative to the current. Our findings indicate that the measured voltage can arise due to fringe-field-induced Hall voltages, different from current-induced spin polarization of the surface state charge carriers, as reported recently. Understanding the nontrivial origin of the measured voltage is important for realizing spintronic devices with topological insulators.  

\end{abstract}

\maketitle

Topological insulators (TIs) belong to a new class of materials with properties dictated by the topology of electronic band structures~\cite{hasan_colloquium:_2010, ando_topological_2013, moore_birth_2010}. These materials have a high spin-orbit coupling which leads to metallic surface states due to parity change in occupied electronic states. The conservation of time-reversal symmetry in the surface states gives rise to the property of spin-momentum locking, where spin and momentum of the surface state charge carriers are directly related in a perpendicularly right-handed orientation [Fig.~\ref{fig:MeasurementGeometryAndResultsV1}(a)]. This spin texture can be compared with that originating from Rashba spin-orbit coupling, but consists of a single Fermi circle with opposite spin texture. This direct coupling between spin and momentum in TIs should allow electrical injection and detection of spin currents in spintronic structures without the need of ferromagnetic layers. The time-reversal symmetry also leads to robustness against elastic backscattering from non-magnetic impurities reducing the surface conduction dissipation. Furthermore, the surface states are located within the bulk band gap such that these can be addressed independently from the spin-unpolarized bulk. With this combination of properties, topological insulators provide a new platform for spintronics. 

In our study, we use the canonical topological insulator Bi$_2$Se$_3$ which has a bulk band gap of 0.3 eV, making it suitable for potential applications at room temperature. The spin-momentum locking of the surface states of Bi$_2$Se$_3$ has been well investigated by ARPES measurements and 100 $\%$ spin polarization has been reported~\cite{xia_observation_2009, hsieh_tunable_2009, pan_electronic_2011}. Additionally, the existence of spin-momentum locking over a large temperature range has been investigated through measurements using spin transfer torque, inverse spin Hall effect, and tunnel junction devices on Bi$_2$Se$_3$~\cite{mellnik_spin-transfer_2014, deorani_observation_2014, liu_spin-polarized_2015}. Furthermore, experiments on electrical detection of this spin texture have been recently reported in which the texture was analyzed using the three-terminal potentiometric method as discussed in Ref.~\citen{hong_modeling_2012}. It was shown that by using ferromagnetic (FM) contacts, spin polarization in the TI channel can be measured by probing the potential at that contact~\cite{liu_spin-polarized_2015, li_electrical_2014, dankert_room_2014, tang_electrical_2014, tian_electrical_2015, tian_topological_2014, ando_electrical_2014}. For several TI compounds, it has been claimed that the observed change in voltage, upon inverting the ferromagnet's magnetization, originates from the current-induced spin polarization in the surface states. 

Here, we report on our observations of electrically probing spin-momentum locking on well-characterized thin films of Bi$_2$Se$_3$, using ferromagnets to detect the spin polarization for different magnetization directions with respect to the current bias. For this, we use a Hall bar patterned device and change the polarity of the fixed current bias in the TI channel as well as the magnetization of the FM detectors to obtain similar loops in the measured voltage, as reported earlier~\cite{liu_spin-polarized_2015, li_electrical_2014, dankert_room_2014, tang_electrical_2014, tian_electrical_2015}. Additional design flexibility in our device geometry enables the decoupling of spin and current paths, offering a unique possibility to investigate additional magnetoresistance effects to the observed voltage in our TI channel. Surprisingly, we find that the measured voltages for a current bias applied both along and perpendicular to the TI channel exhibit features that cannot be ascribed to the spin polarization of the surface states alone. Instead, similar signals can result from Hall voltages generated by the fringe fields from the FM detector in close proximity with the TI channel.

In this study, we used thin films of Bi$_2$Se$_3$ of 20 nm which were grown by Molecular Beam Epitaxy (MBE) on Al$_2$O$_3$(0001) substrates in a custom-designed SVTA MOS-V-2 MBE system at a base pressure lower than 5$\times$10$^{-10}$ Torr. Bi and Se fluxes were provided by Knudsen cells, and the fluxes were measured using a quartz crystal microbalance~\cite{bansal_thickness-independent_2012}.
\begin{figure}[t]
	\centering
		\includegraphics[width=\columnwidth]{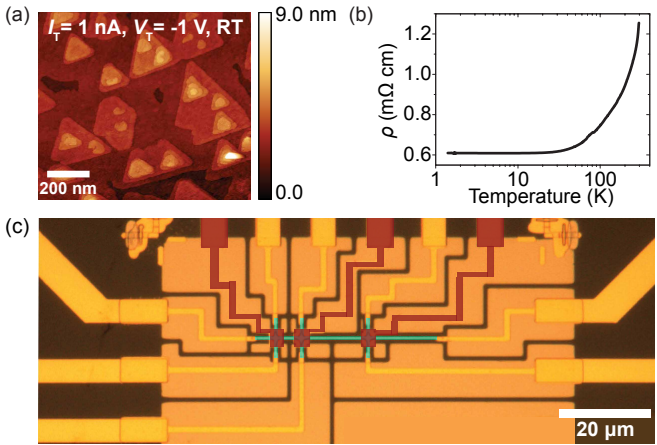}
	\caption{(a) STM topography image taken at room temperature (RT) showing triangular growth spirals ($I_\mathrm{T}$ = 1 nA, $V_\mathrm{T}$ = 1 V). (b) Resistivity $\rho$ versus temperature measurement for a 20 QL Bi$_2$Se$_3$ film. (c) Device structure with Ti/Au contacts (bright yellow) on the Bi$_2$Se$_3$ channel (green), etched structure (dark) and FM contacts (brown).}
	\label{fig:BasicPropertiesAndDeviceStructure}
\end{figure}
A basic investigation with Scanning Tunneling Microscopy (STM) shows triangular growth spirals particular for the Bi$_2$Se$_3$ crystal growth mode [Fig.~\ref{fig:BasicPropertiesAndDeviceStructure}(a)]. Cross-sectional measurements on these triangles reveal a step height of 1 nm corresponding to the height of one unit cell of the crystal [quintuple layer (QL)]. A typical resistivity $\rho$ versus temperature plot for this film [Fig.~\ref{fig:BasicPropertiesAndDeviceStructure}(b)] shows a decrease in resistivity upon decreasing temperature, indicating electron-phonon scattering to be the dominant scattering mechanism. Combining this data with standard Hall measurements yields a temperature-independent bulk charge carrier density of 1.25$\times$10$^{19} $cm$^{-3}$ and a mobility of 385 and 815 cm$^2$(Vs)$^{-1}$ at 300 K and 1.5 K, respectively. 

The advantage of using Bi$_2$Se$_3$ thin films over single crystals is its design flexibility that allows for the investigation of parasitic effects on the measured voltage. In this work, the devices were first patterned with Ohmic contacts using a combination of deep-UV lithography and Electron Beam Lithography (EBL) techniques. The Ohmic contacts consist of 5 nm Ti/70 nm Au deposited by electron beam evaporation. Thereafter, Hall bar structures with a channel width of 1 $\micro$m were realized using EBL and Ar plasma dry etching. In the last step, spin detector contacts on the TI channel were fabricated by growing tunnel barriers of 2 nm TiO$_2$ deposited by evaporating Ti, followed by \textit{in situ} oxidation in an O$_2$ atmosphere of 300 mTorr. The FM layer is a 35 nm Co layer and is capped by 5 nm of Au. The ferromagnetic contacts have typical lateral dimensions of 1-3 $\micro$m. The values for the resistance-area product ($RA$) of the spin contacts are in the range of 3-100 k$\Omega\micro$m$^2$. The resulting device structure is as shown in Fig.~\ref{fig:BasicPropertiesAndDeviceStructure}(c).

In our measurements, a current $I_\mathrm{e}$, that indicates the direction of flow of the electrons, is sent through the TI channel along the $x$ direction such that an imbalance in the momentum is created. Due to spin-momentum locking, this imbalance in momentum leads to a net spin polarization of the surface state charge carriers \textbf{$\sigma$} perpendicular to $I_\mathrm{e}$ as shown in Fig.~\ref{fig:MeasurementGeometryAndResultsV1}(a). The ferromagnetic contact with magnetization \textbf{M} is used to probe voltages $V_\mathrm{1}$ and $V_\mathrm{2}$ in the TI channel with respect to an Ohmic contact designed outside the current path to minimize charge-related effects. In this geometry, there is no net charge current flowing through the detector such that only the (spin) potential at the channel is measured and barrier effects do not play a role. The magnetization of the FM contact, and therefore the spin sensitivity, can be inverted by application of a magnetic field $B$ along the $y$-axis. The measurements are done for three different geometries, labeled A, B and C, in which the relative orientation between the spin polarization and magnetization is varied. The measurements are performed in a flow cryostat system with magnetic fields up to 1 T, using ac modulation techniques. The results presented here are representative of multiple devices. 

\begin{figure}[t]
	\centering
		\includegraphics[width=\columnwidth]{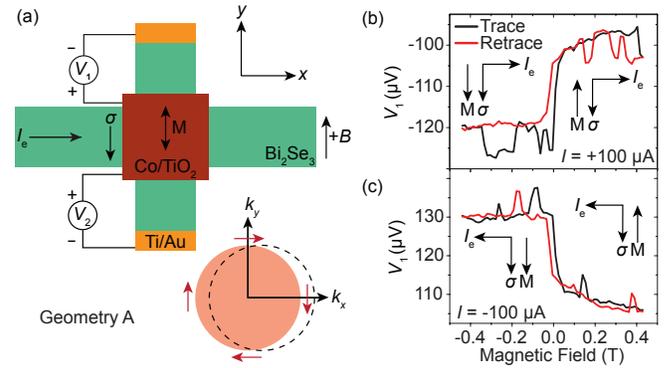}
	\caption{(a) Measurement geometry A in which measured voltages $V_\mathrm{1}$ and $V_\mathrm{2}$ upon application of a current $I_\mathrm{e}$ along the $+x$ direction are indicated. (b) Voltage $V_\mathrm{1}$ versus magnetic field $B$ using the conventions drawn in (a) for a current $I_\mathrm{e}$ in the $+x$ direction at a bias of 100 $\micro$A. Measured at $T$ = 4 K. (c) Same measurement but at an inverted bias of -100 $\micro$A. The insets indicate the different orientations of spin polarization $\sigma$, magnetization \textbf{M} and the direction of flow of the electrons $I_\mathrm{e}$. }
	\label{fig:MeasurementGeometryAndResultsV1}
\end{figure}

In the first measurement geometry, labeled as geometry A, the voltage between the ferromagnetic contact ($RA$ = 3 k$\Omega\micro$m$^2$) and the Ohmic contacts is measured at 4 K sourcing an ac current bias of +100 $\micro$A. The voltage is recorded while sweeping a magnetic field $B$ from positive (in the $+y$ direction) to negative ($-y$; indicated by trace) and back (retrace) as shown in Fig.~\ref{fig:MeasurementGeometryAndResultsV1}. Voltage $V_\mathrm{1}$ shows a clear switch at $\sim$15 mT due to magnetization inversion of the Co contacts. The voltage difference $\Delta V$ between the magnetization directions is indicative of a spin polarization in the TI channel, as reported in earlier works~\cite{liu_spin-polarized_2015, li_electrical_2014, dankert_room_2014, tang_electrical_2014, tian_electrical_2015, tian_topological_2014, ando_electrical_2014}. This polarization can be explained by the spin-momentum locking of the surface state charge carriers, as indicated by the different orientations of the magnetization \textbf{M} and the surface state's spin polarization \textbf{$\sigma$} [inset Fig.~\ref{fig:MeasurementGeometryAndResultsV1}(b)]. At large positive magnetic fields, the magnetization \textbf{M} is aligned parallel with the spin polarization \textbf{$\sigma$} yielding a larger voltage~\cite{jonker_spin_2004} than in the antiparallel state where the Co magnetization is antiparallel relative to \textbf{$\sigma$}. Upon inverting the current bias to -100 $\micro$A, thereby inverting the spin polarization in the channel, an opposite switch in the voltage is obtained, indicating that the spin polarization in the channel is indeed reversed in accordance with the surface state spin texture [Fig.~\ref{fig:MeasurementGeometryAndResultsV1}(c)]. The background signal that is observed in the measurements originates from charge-related effects and therefore does not contribute to magnetization-dependent effects in the measured voltage. The additional small jumps in the signal are probably due to instabilities of the FM layer and scale with bias. Furthermore, magnetization-dependent voltage signals were measured for FM contacts with $RA$ = 19 k$\Omega\micro$m$^2$ (discussed in Supplemental Material~\cite{Supplementary_PRBRapCom_2015}), which show similar features as in Fig.~\ref{fig:MeasurementGeometryAndResultsV1} and therefore magnetoresistance effects can be excluded.

To rule out any artifacts related to the geometry of the contacts, voltage $V_\mathrm{2}$ was measured with respect to the opposite Ohmic contact (Fig. S2 in Supplementary Material~\cite{Supplementary_PRBRapCom_2015}). This voltage shows a similar switching behavior but with a different magnitude of $\Delta V$ and background signal, while the Ohmic contact properties are the same. Subtracting $V_\mathrm{1}$ from $V_\mathrm{2}$ (shown in Fig. S2~\cite{Supplementary_PRBRapCom_2015}) reveals a residual switching behavior which indicates that the voltage at the Ohmic contacts is affected by the magnetization switching of the FM contacts. The difference between $V_\mathrm{1}$ and $V_\mathrm{2}$ might be due to a slight asymmetry in the design of the FM contact on both sides of the Hall bar. Any extracted $\Delta V$ is thus very sensitive to the relative alignment of the contacts. 

Assuming that the voltage difference $\Delta V$ arises due to current-induced spin polarization of the surface state charge carriers $P_\mathrm{SS}$, one can write (see Supplemental Material for details~\cite{Supplementary_PRBRapCom_2015}):
\begin{equation}
\label{eq:SurfaceStatePolarization}
{\Delta}V={\eta}P_\mathrm{FM}P_\mathrm{SS}\frac{2h}{e^2}\frac{2\pi}{k_\mathrm{F}W}I_\mathrm{e},
\end{equation}

where $\eta$ is the thickness ratio $t_\mathrm{SS}/t_\mathrm{Total}$ that accounts for parallel bulk conduction (assuming conductivities for bulk and surface are the same), $P_\mathrm{FM}$ is the spin polarization of the ferromagnetic contact, $k_\mathrm{F}$ the Fermi wave vector, $W$ the channel width and $I_\mathrm{e}$ the applied current bias. With $k_\mathrm{F}$ = 0.072 {\AA}$^{-1}$ calculated from the bulk charge carrier density, $W$= 1 $\micro$m, a fixed $\eta$ = 0.1 (extent of the surface states is taken to be 2 QL~\cite{yazyev_spin_2010}) and $P_\mathrm{FM}$ ranging from 3 to 30 $\%$, we obtain values for the surface state spin polarization of $P_\mathrm{SS}$ ranging from 1.5 to 15 $\%$. Values of $P_\mathrm{SS}$ $\approx$ 15 $\%$ for low $P_\mathrm{FM}$ are comparable to the values previously reported for Bi$_2$Se$_3$~\cite{liu_spin-polarized_2015, li_electrical_2014, dankert_room_2014, tian_topological_2014}, whereas $P_\mathrm{SS}$ $\approx$ 1.5 $\%$ (high $P_\mathrm{FM}$) is in the range of those extracted for the counterdoped compounds~\cite{tang_electrical_2014, tian_electrical_2015, ando_electrical_2014} but are always lower than the theoretical limit for electrical transport~\cite{yazyev_spin_2010}. Similar values of $P_\mathrm{SS}$ have been obtained for contacts with higher $RA$ showing a slight increase in $\Delta V$ (see Supplemental Material~\cite{Supplementary_PRBRapCom_2015}). Contributions from bulk states or from other parallel surface state channels can change the value of $P_\mathrm{SS}$ significantly~\cite{bianchi_coexistence_2010}. 

Our device design offers the flexibility to source the current along the $y$ axis and measure the voltage $V_\mathrm{1}$ and $V_\mathrm{2}$ between the FM contact and lateral Ohmic contacts as shown in Fig.~\ref{fig:ResultsAlternativeGeometry}(a) (labeled as geometry B). Such an investigation allows us to know more about the origin of the measured voltage in Fig.~\ref{fig:MeasurementGeometryAndResultsV1} which is attributed to spin-momentum locking in earlier reports~\cite{liu_spin-polarized_2015, li_electrical_2014, dankert_room_2014, tang_electrical_2014, tian_electrical_2015, tian_topological_2014, ando_electrical_2014}. In this measurement geometry, the spin polarization of the surface state charge carriers is oriented perpendicular to the magnetization of the Co contacts, implying that these contacts should not detect any spin polarization in the TI channel. Surprisingly, we see a clear change in $V_\mathrm{1}$ and $V_\mathrm{2}$ for this measurement geometry too [Fig.~\ref{fig:ResultsAlternativeGeometry}(b) and~\ref{fig:ResultsAlternativeGeometry}(c)].  

\begin{figure}[t]
	\centering
		\includegraphics[width=\columnwidth]{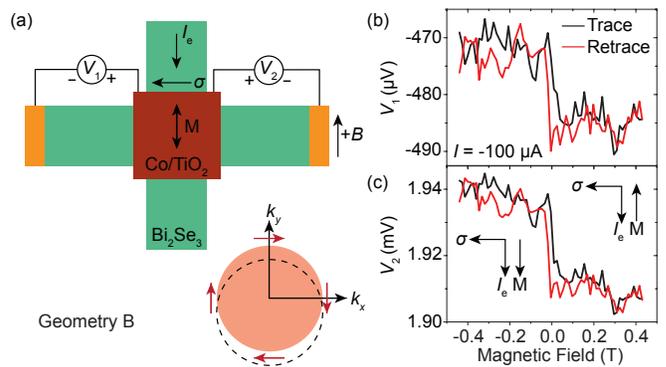}
	\caption{(a) Alternative geometry B in which the current is biased along the $y$ axis and voltages $V_\mathrm{1}$ and $V_\mathrm{2}$ are measured in the horizontal extent at both sides of the FM contact. (b) and (c) Measurements of respectively $V_\mathrm{1}$ and $V_\mathrm{2}$ versus magnetic field $B$  at a bias of -100 $\micro$A and at $T$ = 4 K. The insets indicate the different orientations of spin polarization $\sigma$, magnetization \textbf{M} and the direction of flow of the electrons $I_\mathrm{e}$.} 
	\label{fig:ResultsAlternativeGeometry}
\end{figure}

\begin{figure}[b]
	\centering
		\includegraphics[width=\columnwidth]{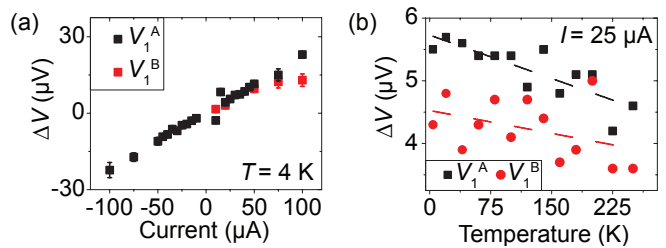}
	\caption{(a) Current bias dependence on the voltage difference $\Delta V$ for $V_\mathrm{1}$ in geometry A and geometry B at $T$ = 4 K. (b) Temperature dependence of $\Delta V$ for $V_\mathrm{1}$ in geometries A and B at $I$ = 25 $\micro$A. Dashed lines are a guide to the eye.}
	\label{fig:BiasDependenceAndTemperatureDependence}
\end{figure}
Upon investigating the bias dependence of $\Delta V$ with applied current for geometry B [Fig.~\ref{fig:BiasDependenceAndTemperatureDependence}(a)], it is observed that the extracted values are on the same order of magnitude as that for the original geometry A. The voltage difference $\Delta V$ scales linearly with current as can be expected from Eq.~\ref{eq:SurfaceStatePolarization}. The background voltage also scales linearly with current bias (not shown here) as expected from the Ohmic background. The similarities in the magnitude and trend of the measured voltages in both geometries A and B indicates a different origin than due to spin-momentum locking. The observed isotropy of the measured voltages for different current directions also rules out its origin due to hexagonal warping in the TI~\cite{wang_observation_2011,alpichshev_stm_2010} or effects related to a tilted magnetization of the FM detector.

Furthermore, for the FM contact with $RA$ = 3 k$\Omega\micro$m$^2$, the temperature dependence of $\Delta V$ for geometries A and B has been plotted in Fig.~\ref{fig:BiasDependenceAndTemperatureDependence}(b). We find a weak temperature dependence of $\Delta V$ up to 250 K as also reported in Ref.~\citen{dankert_room_2014}. If we assume $\Delta V$ to arise due to surface states in the TI, we can infer that the surface state spin polarization does not change appreciably up to 250 K while beyond 250 K its detection is mostly limited due to a change in the contact's properties. The small decrease in $\Delta V$ for both geometries cannot be explained fully by the temperature-dependent resistivity of Bi$_2$Se$_3$ [Fig.~\ref{fig:BasicPropertiesAndDeviceStructure}(b)] and suggests a possible decrease in detection efficiency of the FM contact with increasing temperature. The wide temperature range over which the generated spin polarization can be detected is surprising when compared to reports on the counterdoped compounds, where signals disappear at low temperatures~\cite{tang_electrical_2014, tian_electrical_2015, ando_electrical_2014}.

In yet another measurement geometry (geometry C; Fig.~\ref{fig:ResultsAlternativeGeometryII}), the sample is aligned with respect to the magnetic field such that the magnetization is directed along the channel. In this measurement geometry, where the magnetization of the FM is rotated by 90$^\circ$ as compared to geometry A, the FM contacts should not detect a spin polarization in the TI channel. However, we do observe a clear voltage difference $\Delta V$, as shown in Figs.~\ref{fig:ResultsAlternativeGeometryII}(b) and~\ref{fig:ResultsAlternativeGeometryII}(c), when the magnetization of the FM is switched. This differs from that reported by Li \textit{et al.}~\cite{li_electrical_2014}. The linear background as observed in Fig.~\ref{fig:ResultsAlternativeGeometryII} is due to an unintended misalignment of the device with respect to the external magnetic field. This results in an out-of-plane field component leading to a Hall voltage (for a misalignment of 0.5$^{\circ}$, we find this to be 9 $\micro$V in good agreement with the slope in the voltage measured at 0.4 T).

\begin{figure}[t]
	\centering
		\includegraphics[width=\columnwidth]{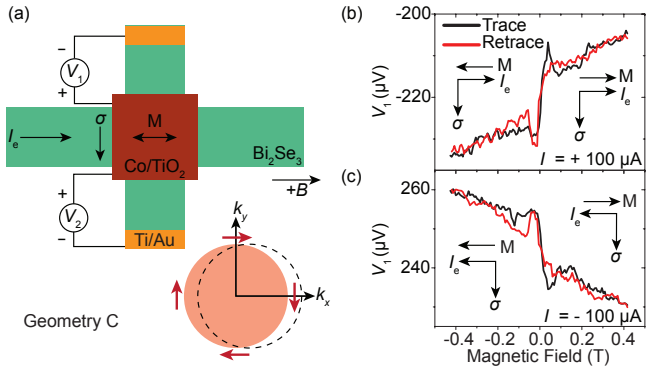}
	\caption{(a) Alternative geometry C where the current is biased along the $x$ axis and voltages $V_\mathrm{1}$ and $V_\mathrm{2}$ are measured in the vertical direction with a rotated magnetization of the detector contact. (b) Measurement of $V_\mathrm{1}$ versus magnetic field $B$ at a bias of +100 $\micro$A and $T$ = 4 K. (c) Same measurement but at an inverted bias of -100 $\micro$A and $T$ = 4 K. The insets indicate the different orientations of spin polarization $\sigma$, magnetization \textbf{M} and the direction of flow of the electrons $I_\mathrm{e}$.}
	\label{fig:ResultsAlternativeGeometryII}
\end{figure}

The similarities in the observed signals for all different measurement configurations raise questions on their origin. Rashba spin-orbit coupling or spin Hall effects can be excluded, since similar spin textures (momentum perpendicular to the spin orientation) should not be observed in the alternative geometries B and C. However, the fringe fields arising due to the proximity of the FM layer to the TI channel could mimic a similar voltage for all the different measurement configurations as shown in Figs.~\ref{fig:MeasurementGeometryAndResultsV1},~\ref{fig:ResultsAlternativeGeometry} and~\ref{fig:ResultsAlternativeGeometryII}. To illustrate this, we calculate the magnetic fields in the TI channel for the particular shape of the FM contacts (as in Fig.~\ref{fig:BasicPropertiesAndDeviceStructure}c) which is schematically shown in Fig.~\ref{fig:MagFieldCalculations}(a). In Fig.~\ref{fig:MagFieldCalculations}(b), we find a strong out-of-plane ($B_z$) component at the edges of the channel which is on the order of several 100 mT. Furthermore, we find $B_y$ fields in the \textbf{-M} direction (as displayed in Fig.~\ref{fig:MagFieldCalculations}(a)) on the order of 100 mT. The fringe fields in the $z$ direction are present for all different measurement geometries and will lead to the development of local Hall voltages perpendicular to the current due to small geometrical misalignment of the contacts. As also shown in the Supplemental Material~\cite{Supplementary_PRBRapCom_2015}, we find a considerable current spread in the TI channel in the vicinity of both the Ohmic and FM contacts which could amplify the magnetoresistive effects.

\begin{figure}[b]
	\centering
		\includegraphics[width=\columnwidth]{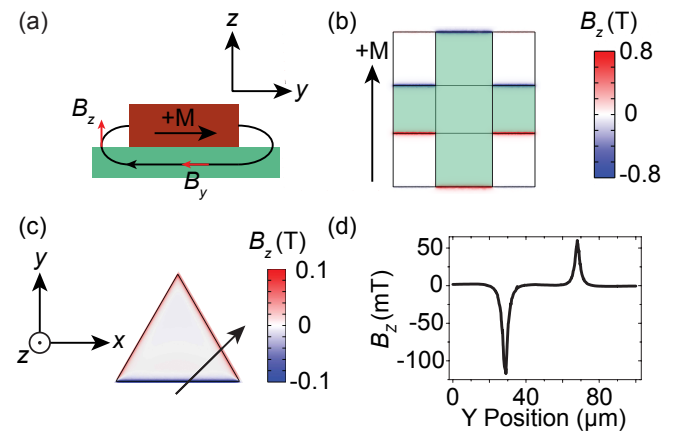}
	\caption{(a) Schematics of fringe fields (black line) arising due to the proximity of ferromagnetic contact with magnetization \textbf{M} (brown) on TI channel (green). Magnetic field components parallel as well as perpendicular to the magnetization are present (indicated in red). (b) Calculation of the out-of-plane component $B_z$ at the interface between FM contact (outer square) and TI channel (inner cross structure; colored false green). At the edges we find magnetic fields around 0.7 T. (c) Calculation of the $B_z$ component at the interface for a triangular step feature with sides of 200 nm typical for the thin films used.(d) Calculated $B_z$ along trace arrow indicated in (c) indicating an asymmetry in the magnetic field.}
	\label{fig:MagFieldCalculations}
\end{figure}

Furthermore, we have also modeled the magnetic field at the interface between the ferromagnetic contact and the channel due to the triangular growth spirals at the Bi$_2$Se$_3$ [Fig.~\ref{fig:BasicPropertiesAndDeviceStructure}(a)]. As shown in Figs.~\ref{fig:MagFieldCalculations}(c) and (d), out-of-plane magnetic fields $B_z$ on the order of 50 mT can develop locally at the edges of the triangular steps. Due to the asymmetry of these triangular features a net magnetic field will act on the charge carriers. This will also give rise to Hall voltages which we calculate to be on the order of several tens of $\micro$V for the device structure used and is thus comparable to the measured voltage signals. Note that in this model we neglect any irregularities in the FM layer which could significantly enhance the fringe fields~\cite{nogaret_electron_2010}. Upon inverting the magnetization of the FM layer, the direction of the fringe fields is inverted which inverts the isotropic local Hall effect and manifests itself as a switch in the measured voltage for all geometries.

In summary, our detailed investigation of the measured voltages in different measurement geometries along with the calculations of the fringe-field-induced voltage make us believe that it is non-trivial to relate the measured voltage to spin polarization of TI surface states alone. Our findings clearly highlight the necessity of a careful analysis of the observed voltage in electrical measurements with topological insulators for future spintronic devices.

The authors would like to thank J. G. Holstein and H. M. de Roosz for the technical support. Work at the University of Groningen is supported by the Dieptestrategie grant from the Zernike Institute for Advanced Materials. Work at Rutgers University is supported by ONR (N000141210456), NSF (DMR-1308142) and Gordon and Betty Moore Foundation's EPiQS Initiative (GBMF4418).

\bibliography{References}
\pagebreak

\newcommand{\beginsupplement}{%
        \setcounter{table}{0}
        \renewcommand{\thetable}{S\arabic{table}}%
        \setcounter{figure}{0}
        \renewcommand{\thefigure}{S\arabic{figure}}%
     }
\newcommand{\diff}{\mathrm{d}}

\newpage
\beginsupplement
\preprint{APS/123-QED}

\maketitle
\onecolumngrid

\section{Derivation of formula for spin voltage $\Delta V$}

To derive an expression for the current induced spin polarization, we start out with the expression for the current density in 3D:

\begin{equation}
\label{eq:Current Density 3D}
\textbf{j}= e\int\frac{\diff\textbf{k}}{(2\pi)^3}\textbf{v}(\textbf{k}),
\end{equation}

which reduces in 2D to:

\begin{equation}
\label{eq:Current Density 2D}
\textbf{j}= e\int\frac{\diff\textbf{k}}{(2\pi)^2}\textbf{v}(\textbf{k}).
\end{equation}

Since we are investigating effects close to the Fermi circle, $\textbf{v}(\textbf{k})$ is constant: ($v_x,v_y$)=($v_\mathrm{F},v_\mathrm{F}$). Upon applying a bias in the $x$-direction, the Fermi circle is shifted with respect to the zero bias position over a length of $\Delta k$. From this, we can define $\diff k_x$ and $\diff k_y$:

\begin{eqnarray}
&\diff k_x=\Delta k\cos\phi, \\
&k_y=k_\mathrm{F}\sin\phi \Rightarrow \diff k_y=k_\mathrm{F}\cos\phi \diff\phi.
\end{eqnarray}

Using these definitions in the expression for the current density, we obtain:

\begin{equation}
\label{eq:Current Density 2D_2}
\textbf{j}_x= \frac{2ev_\mathrm{F}}{(2\pi)^2}\int^{\pi/2}_{-\pi/2}\Delta k k_\mathrm{F}  \cos^2\phi  \diff\phi = \frac{e v_\mathrm{F} \Delta k k_\mathrm{F}}{4\pi},
\end{equation}

where a factor of 2 was included for the contribution to the current density by the removal of states at $-k_x$ and addition of states at $+k_x$. To calculate the associated voltage with the induced current density, the linear dispersion relation is included to the expression for the current density:

\begin{equation}
\label{eq:Linear Dispersion}
E = \hbar k v_\mathrm{F} \Rightarrow= \frac{\diff k}{\diff E} = \frac{1}{\hbar v_\mathrm{F}} \Rightarrow \Delta k = \frac{\Delta E}{\hbar v_\mathrm{F}} \Rightarrow j_x = \frac{e \Delta E k_\mathrm{F}}{2h}.
\end{equation}

The spin orientation revolves around the Fermi circle and to calculate the potential associated with the $x$-projection of the spin, we have:

\begin{equation}
\label{eq:Spin potential}
\mu (\phi) = \Delta E \cos \phi.
\end{equation}

Integrating over angle, one obtains:

\begin{equation}
\label{eq:Spin potential integrated}
qV_{\mathrm{spin}} = \int^{2\pi}_{0} \Delta E \cos\phi \cos\phi \diff\phi = \pi\Delta E.
\end{equation}

Switching the magnetization of the FM detector, changing spin selectivity, would thus lead to a voltage difference of $2\pi\Delta E/e$. Substituting (\ref{eq:Spin potential integrated}) into the expression for the current density (\ref{eq:Linear Dispersion}):

\begin{equation}
\label{eq:Current Density 2D_3}
\textbf{j}_x = \frac{e^2 k_\mathrm{F} \Delta V}{4\pi h}\Rightarrow \Delta V = \frac{h}{e^2} \frac{4\pi}{k_\mathrm{F} W} I.
\end{equation}

In this expression, we assumed pure surface state conduction in the TI channel and 100$\%$ spin polarization of the surface state carriers as well as the magnetization of the FM layer. We correct for this by introducing the ferromagnet's spin polarization $P_\mathrm{FM}$, surface state spin polarization $P_\mathrm{SS}$ and conduction ratio $\eta = n_\mathrm{SS}/n_\mathrm{Total}$.

\section{Additional measurements}

For a FM contact with $RA$ = 19 k$\Omega\micro$m$^2$ with dimensions of 3 x 3.5 $\micro$m$^2$, similar results in the original geometry A as described in the paper are obtained but with higher $\Delta V$ and lower signal-to-noise ratio indicating that the tunnel barrier influences the measurements. For a FM contact with an $RA$-product of 100 k$\Omega\micro$m$^2$, the measured signals were too noisy to observe any switches in the spin voltage. We would like to point out that magnetic doping of the surface states by the FM in the TI channel, thereby breaking time-reversal symmetry, can be excluded since the thick tunnel barrier used in our study, ensures a clear separation of the FM contact from the triangular growth spirals at the TI surface as shown in Fig. 1(a).
\\
\begin{figure}[h]
	\centering
		\includegraphics[width=300pt]{SM_Figure1.pdf}
	\caption{Measurements for $V_\mathrm{1}$ using a contact with $RA$ = 19 k$\Omega\micro$m$^2$ (geometry A).}
	\label{fig:HighRAmeasurement}
\end{figure}

Furthermore, $V_\mathrm{1}$, $V_\mathrm{2}$ and the difference signal $V_\mathrm{2}-V_\mathrm{1}$ similar to those showed in Figure 2 in the paper are plotted for a current bias of 10 $\micro$A in Fig.~\ref{fig:V3 measurement}. One can observe the linear slope in $V_\mathrm{2}$ which is clearly absent in $V_\mathrm{1}$. The difference signal $V_\mathrm{2}-V_\mathrm{1}$ shows a residual switch in the voltage around zero magnetic field, indicating that the fringe field from the ferromagnetic layer affects the potential at the Ohmic contacts too. 

\begin{figure}[t]
	\centering
		\includegraphics[width=\textwidth]{SM_Figure2.pdf}
	\caption{Measurement of $V_\mathrm{1}$, $V_\mathrm{2}$ and the differential signal $V_\mathrm{3}$ (with a zoom in the inset) at a current bias of 10 $\micro$A and $T$ = 4 K.}
	\label{fig:V3 measurement}
\end{figure}

\newpage
\section{Modeling of current spread}

\begin{figure}[h]
	\centering
		\includegraphics[width=\textwidth]{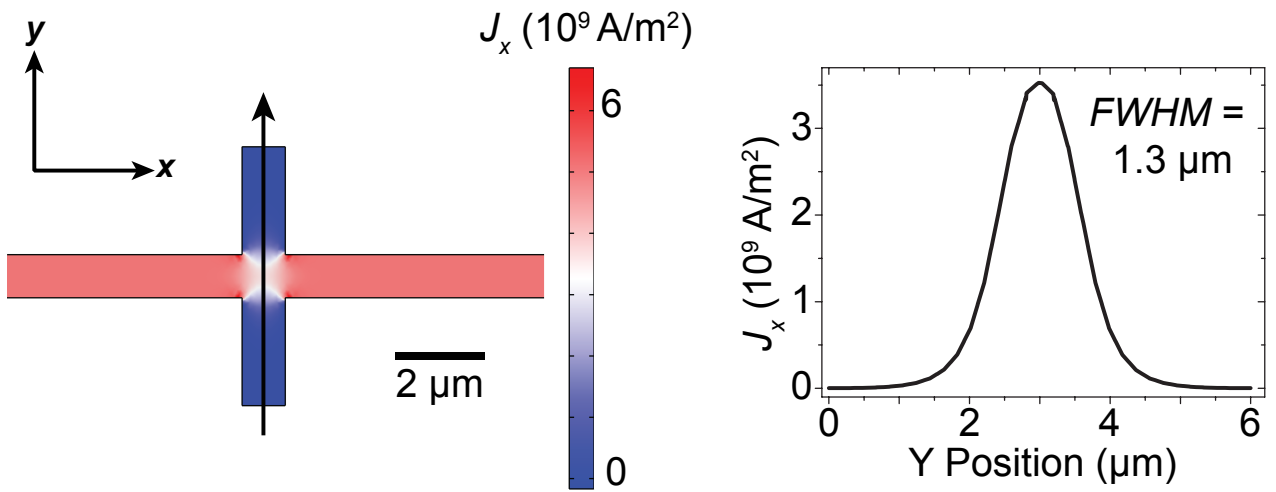}
	\caption{(left) Current density spread at cross-section of Hall bar. (right) Current density along trace arrow indicated in the left schematics. Extracted $FWHM$ is 1.3 $\micro$m. }
	\label{fig:CurrentSpread}
\end{figure}

To estimate the current density at the edges of the FM contact, we modeled the current density $J_x$ around the cross-section, as shown in Fig.~\ref{fig:CurrentSpread}. The extracted $FWHM$ is 1.3 $\micro$m indicating that the current density around edges of the FM (having a width of 3 $\micro$m), where $B_z$ is large, is not negligible and can lead to the occurrence of Hall voltages. Furthermore, we find a considerable current spread in the TI channel in the vicinity of both the Ohmic and FM contacts which might influence the measured voltages as in section S2. Further, the presence of a $J_y$ component experiencing a $B_z$ field will give rise to additional voltages in the $x$-direction. 

\end{document}